\documentclass[review, sort&compress]{elsarticle}

\usepackage{lineno}
\usepackage[colorlinks]{hyperref}
\usepackage[utf8]{inputenc}
\usepackage{enumitem}
\usepackage{graphicx}
\usepackage{amsmath,amssymb,amsfonts}
\usepackage{algorithmic}
\usepackage{textcomp}
\usepackage[table]{xcolor}
\usepackage{gensymb}
\usepackage{float}
\usepackage{csquotes}
\usepackage{mathtools}  
\usepackage{mathrsfs}   
\usepackage{tikz}
\usepackage{mdframed}
\usepackage{caption}
\usepackage{multirow}
\usepackage{placeins}
\usepackage{subcaption}

\AtBeginDocument{
\hypersetup{
urlcolor=blue,
citecolor=green,
linkcolor=red,
}%
}

\modulolinenumbers[1]
\definecolor{bounding_box}{RGB}{77, 121, 255}
\journal{Artificial Intelligence in Medicine}

\begin{document}

\begin{frontmatter}

\title{AG-CUResNeSt: A Novel Method for Colon Polyp Segmentation}

\author[dvs]{Dinh Viet Sang\corref{mycorrespondingauthor}\fnref{equal}}
\ead{sang.dinhviet@hust.edu.vn}

\author[dvs]{Tran Quang Chung\fnref{equal}}
\ead{chung.tqcb190214@sis.hust.edu.vn}

\author[dvs]{Phan Ngoc Lan}
\ead{lan.pn202634m@sis.hust.edu.vn}

\author[2,3]{Dao Viet Hang}
\ead{daoviethang@hmu.edu.vn}

\author[2,3]{Dao Van Long}
\ead{bsdaolong@yahoo.com}

\author[4]{Nguyen Thi Thuy}
\ead{ntthuy@vnua.edu.vn}

\cortext[mycorrespondingauthor]{Corresponding author}
\fntext[equal]{These authors contributed equally to this work}
\address[dvs]{Hanoi University of Science and Technology, Hanoi, Vietnam}
\address[2]{Hanoi Medical University, Hanoi, Vietnam}
\address[3]{The Institute of Gastroenterology and Hepatology, Hanoi, Vietnam}
\address[4]{Faculty of Information Technology, Vietnam National University of Agriculture, Hanoi, Vietnam}



\newcommand*{\DrawBoundingBox}[1][]{%
\draw [red, very thick, #1]
([shift={(-0pt,-0pt)}]current bounding box.south west)
rectangle
([shift={(0pt,+0pt)}]current bounding box.north east);
}

\begin{abstract}
Colorectal cancer is among the most common malignancies and can develop from high-risk colon polyps. Colonoscopy is an effective procedure to detect and remove polyps, especially in the case of precancerous lesions. However, the missing rate in clinical practice is relatively high due to many factors. The procedure could benefit greatly from using AI models for automatic polyp segmentation, which provides valuable insights for improving colon polyp detection. However, precise segmentation is still challenging due to appearance variations of polyps. This paper proposes a novel neural network architecture called AG-CUResNeSt, which enhances Coupled UNets using the robust ResNeSt backbone and attention mechanisms. The network is capable of effectively combining multi-level features and leveraging semantic information flow to yield high accurate polyp segmentation. Experimental results on five popular benchmark datasets show that our proposed method achieves state-of-the-art accuracy compared to existing methods.
\end{abstract}


\end{frontmatter}


\section{Introduction}

Colorectal cancer (CRC) is a leading cause of cancer deaths worldwide, with roughly 694,000 fatalities each year \cite{bernal2017comparative}. Most CRC arises from colon polyps, especially the adenomas with high-grade dysplasia \cite{gschwantler2002high}. According to a longitudinal study \cite{corley2014adenoma}, each 1\% of adenoma detection rate increase is associated with a 3\% decrease in the risk of colon cancer. Therefore, the detection and removal of polyps at the early stage are of great importance to prevent CRC. Nowadays, colonoscopy is considered the gold standard for colon screening and is recommended in many guidelines of different societies \cite{issa2017colorectal}.
Nevertheless, the overloaded healthcare systems in many countries, especially in limited-resource settings, might result in shorter endoscopy duration to guarantee the required number of procedures per day. This factor, combined with low-quality endoscopy systems and experience gaps among endoscopists at different levels of the healthcare system, seriously affects the quality of colonoscopy procedures and increases the risk of missing lesions and inaccurate diagnosis \cite{lee2008adequate, armin2015visibility}. A literature review has shown that the colon polyp missing rate in endoscopies could range from 20-47\% \cite{leufkens2012factors}. Hence, studies to develop computer-aided systems to support endoscopists in providing accurate polyp regions are much needed in both aspects of training endoscopists and application in clinical practice.

In recent years, with the advancement of artificial intelligence (AI), particularly deep learning (DL), we can address the limitations of conventional endoscopy systems. Attempts have been made to build computer-aided diagnosis (CAD) systems for the automatic detection and prediction of polyps, which may help clinicians identify lesions and reduce the miss detection rate  \cite{mesejo2016computer,zhou2019951e,kudo2019artificial}. In some retrospective studies, AI has demonstrated promising results in supporting colon polyp detection \cite{urban2018deep, viscaino2019machine}.
The CAD system is expected to support endoscopists in lesion detection, diagnosis, and quality assurance. It can be applied to solve practical problems, including improving the lesion detection rate, supporting doctors in optimizing strategy during endoscopy for high-risk lesions, and serving the increasing number of patients while maintaining diagnostic quality \cite{chen2018accurate, bisschops2019advanced}.

Despite the growth of many advanced machine learning and computer vision techniques in recent years, automatic polyp segmentation is still a challenging problem. The first challenge is that polyps are often diverse in appearances, such as shape, size, texture, and color. Secondly, as the nature of medical images, the boundary between polyps and their surrounding mucosa, especially in case of flat lesions or unclean bowel preparation, is not always clear during colonoscopy, leading to confusion for segmentation methods.

There are different approaches to polyp segmentation. Traditional machine learning methods are based on hand-crafted features for image representation  \cite{iwahori2013automatic, silva2014toward}. These methods rely on color, texture, shape, or edge information as extracted features and train classifiers to distinguish polyps from surrounding normal mucosa. However, hand-crafted features are limited in representing polyps due to their high intra-class diversity and low inter-class variation between them and hard negative mucosa regions. Recently, deep neural networks have proven to be more effective in solving medical image segmentation problems, particularly those related to endoscopic images of the human gastrointestinal (GI) tract. Among various deep models, encoder-decoder based networks like UNet family \cite{ronneberger2015u} have demonstrated impressive performance. In UNets, high-level semantic features in the decoder are gradually up-sampled and fused with corresponding low-level detailed information in the encoder through skip connections. Inspired by the success of UNets, UNet++ \cite{zhou2019unet++} and ResUNet++ \cite{jha2019resunet++} were proposed for polyp segmentation and yielded promising results. However, these methods heavily depend on the dense concatenation of feature maps at multiple levels, resulting in high computational resource requirements and time-consuming procedures. 

Recently, the attention mechanism has been widely used in various deep learning models, allowing them to focus on learning valuable information from the input. In \cite{oktay2018attention}, Oktay et al. introduce attention gates to UNets in order to suppress irrelevant low-level information from encoders before concatenating them with high-level feature maps in decoders. Fan et al. \cite{fan2020pranet} enhanced an FCN-like model by a parallel partial decoder and reverse attention module and obtained impressive results.
On the other hand, previous works show that stacking multiple UNets allows the networks to learn a better feature representation and considerably improves the accuracy. DoubleUNet \cite{jha2020doubleu} stacks two UNets on top of each other, and it was applied for polyp segmentation. However, DoubleUNet lacks skip connections between the two UNets, which limits the information flow within the network. Another weakness of DoubleUNet is the use of an old VGG-19 backbone, which can be replaced with more efficient models proposed recently, e.g., the ResNet family \cite{xie2017aggregated,he2016deep,zhang2020resnest}. In \cite{tang2019cu}, Tang et al. introduce the coupled UNets (CUNet) architecture, where coupling connections are utilized to improve the information flow across UNets. In \cite{na2020facial}, Na et al. introduce coupled attention residual UNets and use it as a generator for adversarial learning based Facial UV map completion. The model in \cite{na2020facial} is based on the ResNet backbone and utilizes the fast normalized fusion \cite{tan2020efficientdet} to combine the information between the two UNets, which can result in potential information loss.

In this paper, we propose a new deep network, named AG-CUResNeSt, that addresses the above limitations for efficient polyp segmentation. Our main contributions are:
\begin{itemize}
\item A novel network architecture based on coupled UNets with improved mechanisms for integrations of skip connections and attentions. In which, the encoders are strengthened by residual connections and split-attention blocks. 
Attention gates are integrated into skip connections within each UNet to suppress the redundant low-level information from the encoders. Skip connections across the two UNets 
are leveraged to reduce gradient vanishing and promote feature reuse.
\item An extensive set of experiments on popular benchmark datasets shows that our method yields superior accuracy than other state-of-the-art approaches. Especially, our approach achieves significantly better cross-dataset generalization than others when all models are trained on one dataset and tested on another dataset.
\end{itemize}

The rest of the paper is structured as follows. Section~\ref{sec:related} reviews the literature regarding CNN backbones and semantic segmentation in medical image analysis. In Section~\ref{sec:propose}, we describe the proposed network architecture in detail. Section~\ref{sec:experiments} outlines our experiment settings. The results are presented and discussed in Section \ref{sec:results}. Finally, we conclude the paper and outline future works in Section~\ref{sec:conclude}.

\section{Related Work}
\label{sec:related}
There have been many methods proposed for semantic image segmentation in general and for the purpose of segmenting the polyps in colonoscopy images in particular. This section briefly reviews prior methods related to our work, focusing on deep neural network models for colorectal polyps segmentation.

\subsection{CNN Architectures}
Since AlexNet \cite{krizhevsky2012imagenet}, Convolutional Neural Networks (CNNs) have dominated in solving computer vision tasks. VGG \cite{simonyan2014very} proposes a simple yet effective modularized network design exploiting the efficiency of small $3 \times 3$ kernels. However, plain networks like VGG suffer from degradation when their depth increases. ResNet \cite{he2016deep} introduces identity skip connections to smooth out the objective function's landscape. Skip connections also reduce gradient vanishing and allow very deep networks to learn better feature representations. GoogleNet \cite{szegedy2015going} demonstrates the success of multi-branch networks, where each branch is carefully designed using different convolutional kernel sizes. ResNeXt \cite{xie2017aggregated} improves ResNet with a unified multi-branch design, where all branches have the same architecture. SE-Net \cite{hu2018squeeze} introduces a channel attention mechanism that adaptively recalibrates channel-wise feature responses. SK-Net \cite{li2019selective} proposes an adaptive selection mechanism to fuse two network branches to adaptively adjust receptive field sizes of neurons according to the input. Recently, ResNeSt \cite{zhang2020resnest} integrates the channel-wise attention on different network branches to exploit their success in capturing cross-feature interactions and learning diverse representations. Besides, with the growth of computing capability, some efficient CNNs such as EfficientNet \cite{tan2019efficientnet} are automatically designed by machines thanks to neural architecture search techniques.

\subsection{Semantic Segmentation for Medical Image Analysis}
Semantic image segmentation has been an area of very active research in recent years. Many network architectures and learning techniques have been proposed to improve segmentation accuracy, latency, and throughput. In \cite{long2015fully}, the authors utilize several well-known classification networks (AlexNet \cite{krizhevsky2012imagenet}, VGG \cite{simonyan2014very} and GoogLeNet \cite{szegedy2015going}) in segmentation networks, coupled with transfer learning techniques. UNet \cite{ronneberger2015u} is among the most famous network architectures for segmentation. The network consists of an encoder and a decoder, with skip connections between corresponding levels. More recently, DeepLabV3 \cite{chen2017rethinking} introduces atrous convolutions to extract denser features for better performance in segmentation tasks.

Semantic segmentation specifically for medical images has also attracted much attention. UNet is among the first successful applications of neural architecture in medical images and brings several variants in the following years. In \cite{zhou2019unet++}, Zhou et al. introduce UNet++, an ensemble of nested UNets of varying depths, which partially share an encoder and jointly learn using deep supervision. Later, Jha et al. \cite{jha2019resunet++} propose ResUNet++ that takes the advantages of residual blocks, squeeze and excitation units, atrous spatial pyramidal pooling (ASPP), and the attention mechanism.

DoubleUNet \cite{jha2020doubleu} stacks two UNet blocks with a pre-trained VGG backbone, exploits squeeze and excitation units, and ASPP modules. The performance of this network surpasses previous methods on several different datasets. However,  DoubleUNet suffers from using the old VGG backbone and the lack of skip connections across the two UNet blocks, limiting the information flow. CUNet \cite{tang2019cu} improves the information flow by adding skip connections across UNet blocks. Attention UNet \cite{oktay2018attention} introduces attention gates, which suppress unimportant regions while highlighting salient features useful for a specific task.

Non-UNet architectures are also utilized for medical segmentation. Fan et al. \cite{fan2020pranet} propose PraNet, enhancing an FCN-like model using a parallel partial decoder and reverse attention modules. PraNet achieves state-of-the-art performance on five challenging benchmark medical datasets.  HarDNet-MSEG \cite{huang2021hardnet} presented a slightly similar architecture as PraNet and optimized for inference speed. Jha et al. \cite{jha2021real} recently proposed ColonSegNet, a very lightweight architecture that achieved highly competitive performance on the Kvasir-SEG dataset.

In this paper, we analyze and integrate the three aforementioned architectures  \cite{tang2019cu,jha2020doubleu,oktay2018attention} to propose a novel network that outperforms previous methods.

\section{Method}
\label{sec:propose}
\begin{figure*}[ht!]
\includegraphics[width=350pt]{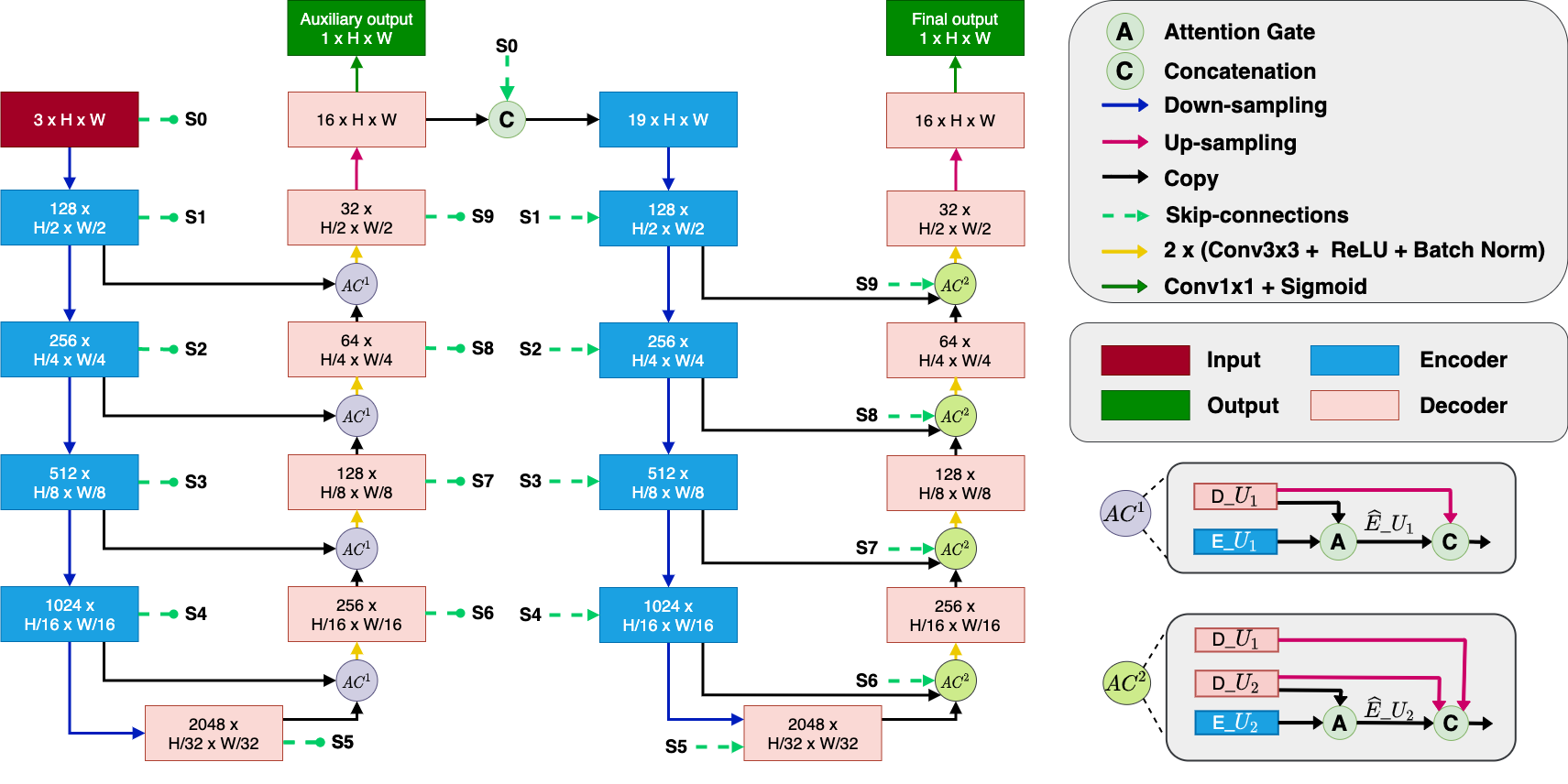}
\caption{An overview of the proposed AG-CUResNeSt. Attention gates within each UNet are used to suppress irrelevant information in the encoder's feature maps. Skip connections across the two UNets are also utilized to boost the information flow and promote feature reuse.}
\label{fig:fig_attention_ResCUNeSt}
\end{figure*}

\subsection{Overall Architecture}
Stacking multiple UNet-like blocks has proven to be efficient in many previous works \cite{newell2016stacked, jha2020doubleu, tang2019cu}. Inspired by this idea, we design a network that consists of two coupled UNets with similar architectures. The overall architecture of our proposed network is depicted in Fig \ref{fig:fig_attention_ResCUNeSt}. Each UNet has an encoder and a decoder with skip connections between them. The encoder takes an image input of size $512 \times 512$, then passes through five top-down blocks to produce a high-level semantic feature map of size $16 \times 16$. This feature map is gradually up-sampled through five bottom-up blocks of the decoder and fused with low-level information in the encoder via gated skip connections called attention gates. The role of attention gates is to suppress the irrelevant information from the encoder before it is concatenated with the decoder. This helps the network focus better on the important regions to yield a good prediction. Next, we use a 1x1 conv layer followed by a sigmoid layer at the end of each UNet to yield its output. The last feature map of the first UNet is combined with the raw input image and then fed to the second UNet for further refinement. Inspired by \cite{tang2019cu}, we also use skip connections across the two UNets to enhance the information flow and promote feature reuse in the network. The skip connections between the two encoders help the second UNet maintain the detailed information in the input image, while those between the two decoders amplify the semantic representation of the second decoder. The skip connections are expected to improve the gradient flow during in backpropagation phase during training. They also smoothen the landscape of the loss function \cite{DBLP:conf/nips/Li0TSG18} and thus facilitate the optimization process.

\subsection{Backbone: ResNet Family}
Plain networks tend to decrease performance on both training and test datasets as their depth increases. This is a widely observed phenomenon called the degradation problem. ResNet \cite{he2016deep} addresses this problem by introducing skip connections.
Suppose that $H(x)$ as an underlying mapping to be fitted by a block of a few nonlinear layers, where $x$ is the input to the first layer. Instead of directly approximating $H(x)$, we can let the block approximate the corresponding residual mapping $F(x) = H(x) - x$. The original mapping can be obtained using a skip connection as $H(x) = F(x) + x$. By this trick, if the network wants to learn identity mappings, it just simply needs to drive the weights of the nonlinear layers in the block toward zero. Hence, residual connections facilitate the optimization of the network at almost no cost.

\begin{figure*}[h!]
\centering
\includegraphics[scale=0.6]{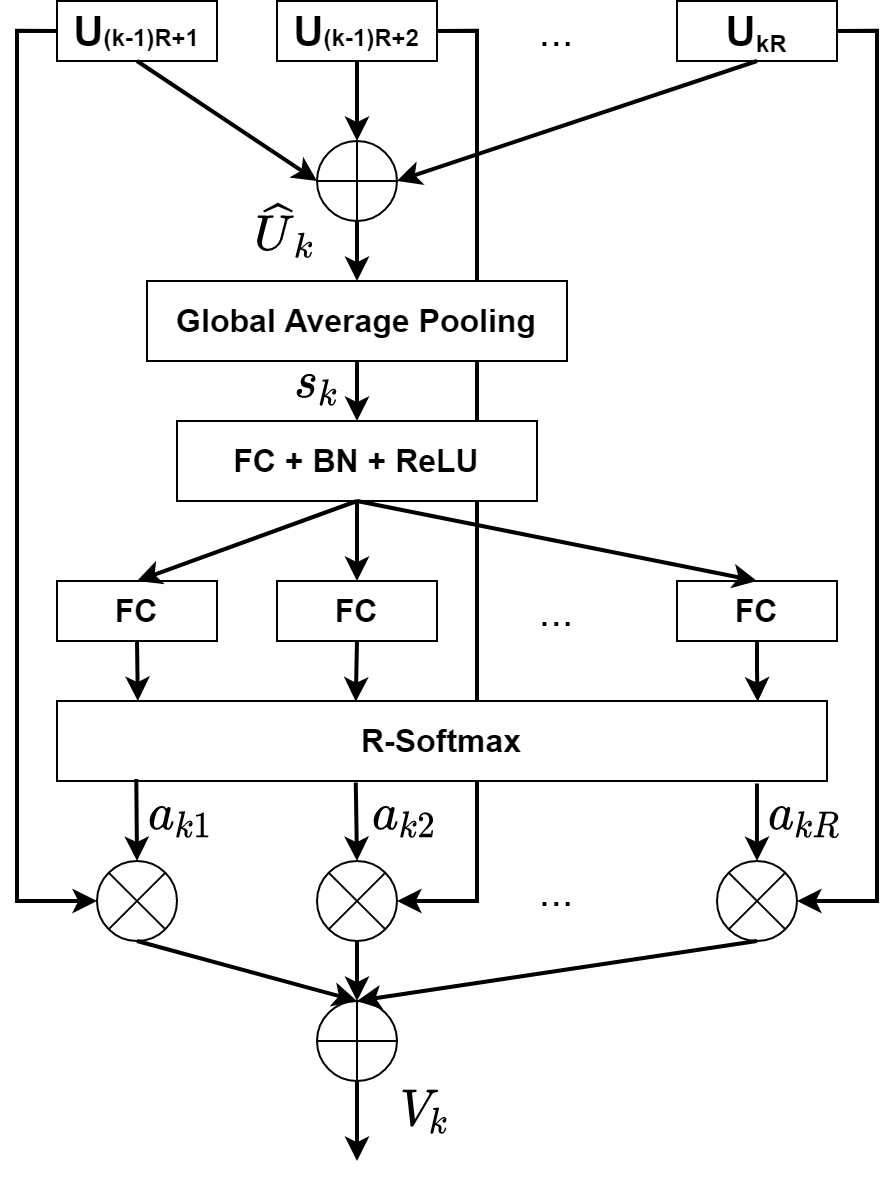}
\caption{Split attention in the $k$-th cardinal group with $R$ splits.}
\label{fig:spit_attention}
\end{figure*}

ResNeXt \cite{xie2017aggregated} introduces a homogeneous multi-branch structure that breaks channel information into $K$ repeated smaller bottleneck branches called cardinal groups.

One of the latest members in the ResNet family is ResNeSt \cite{zhang2020resnest}, which improves the feature representation to boost the performance across multiple computer vision tasks. ResNeSt proposes to split each cardinal group into $R$ smaller feature groups, where $R$ is called a new radix hyperparameter. Hence, the total number of feature groups is $G=K \times R$. Each group is associated with a transformation $\mathcal{F}_i, i = 1,2, \ldots, G$ and outputs an intermediate result $\textbf{U}_i = \mathcal{F}_i(\textbf{X})$, where $\textbf{U}_i \in \mathbb{R}^{H \times W \times C/K}$, and $H, W, C$ are the sizes of a cardinal group's output. The output of $k$-th cardinal group is an element-wise summation of all $R$ splits: ${\hat{\textbf{U}}_k = \sum_{r=1}^{R}\textbf{U}_{R(k-1)+r}}, k = 1, 2, \ldots, K$. Inspired by the ideas of SE-Net \cite{hu2018squeeze} and SK-Net \cite{li2019selective}, ResNeSt introduces the channel-wise attention for multi network splits (Fig.~\ref{fig:spit_attention}). Firstly, the global context information across spatial dimensions $\textbf{s}_k \in \mathbb{R}^{C/K}$ is obtained by applying global average pooling to $\hat{U}_k$. Then a network $\mathcal{G}$ of two consecutive fully connected (FC) layers is added to predict the attention weights over splits in each channel $\textbf{a}_k^c = \{a^c_{k1}, a^c_{k2},...,a^c_{kR}\} \in \mathbb{R}^R$ as follows:

\begin{equation}
a^c_{kr}=\begin{cases}
\frac{exp(\mathcal{G}^c_r(\textbf{s}_k))}{\sum_{j=1}^{R}exp(\mathcal{G}^c_j(\textbf{s}_k))}, \text{if } R > 1,\\
\frac{1}{1+exp(-\mathcal{G}^c_r(\textbf{s}_k))}, \text{ otherwise}.
\end{cases}
\end{equation}

The attention weights corresponding to the $r$-th split can be denoted as $\textbf{a}_{kr} = \{a^1_{kr}, a^2_{kr}, \ldots, a^{C/K}_{kr}\}$.The output of $k$-th cardinal group $\textbf{\textbf{V}}_k \in \mathbb{R}^{H \times W \times C/K}$ is calculated by a weighted fusion over splits:

\begin{equation}
\textbf{V}^c_k = \sum_{r=1}^{R}a^c_{kr}\textbf{U}^c_{R(k-1)+r}, {c = 1, 2, \dots, C/K.}
\end{equation}

Next, the representation of all cardinal groups are concatenated along the channel dimension: $\textbf{V} = Concat(\textbf{V}_1,\textbf{V}_2,\ldots,\textbf{V}_K)$. Finally, a standard skip connection is applied $\textbf{Y} = \textbf{X} + \mathcal{T}(\textbf{X})$, where $\mathcal{T}(\textbf{X})$ is an appropriate transform to align the output shapes if needed.

The experiments in \cite{zhang2020resnest} show that ResNeSt even outperforms the machine designed architecture EfficientNet \cite{tan2019efficientnet} in accuracy and latency trade-off. In this study, we conduct an ablation study on different backbones including ResNet-50, ResNet-101, ResNeSt-50 and ResNeSt-101. Besides, we also compare the ResNet family with Non-ResNet architectures such as VGG and EfficientNet.

\subsection{Attention Gate}
\begin{figure*}[ht!]
\includegraphics[width=350pt]{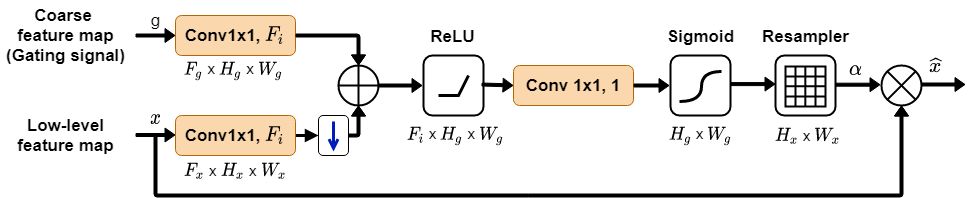}
\caption{The Attention Gate (AG) receives two inputs: a low-level feature map $x$ from an encoder and a coarse feature map $g$ from a corresponding decoder. The feature map $x$ is firstly down-sampled and fused with $g$, then fed to some hidden layers to yield an attention coefficient map $\alpha$. Finally, the input features $x$ are scaled with attention coefficients to suppress irrelevant information.}
\label{fig:attention_gate}
\end{figure*}

Attention gates (AG) \cite{oktay2018attention} can implicitly learn to suppress the irrelevant information in an input image while strengthening salient features necessary for a specific task. In the encoder of a UNet, input data is gradually down-sampled and transformed from low-level to high-level semantic feature maps with coarser scales. In the decoder, coarse feature maps are upsampled and fused with low-level ones to produce a final segmentation result. Fig.~\ref{fig:attention_gate} describes how the attention gate works. Suppose that $g \in \mathbb{R}^{F_g \times H_g \times W_g}$ is a coarse feature map in the decoder that provides information to suppress the irrelevant content in a low-level feature map $x \in \mathbb{R}^{F_x \times H_x \times W_x}$ from the encoder. Each feature map $g$ and $x$ is fed to a $1 \times 1$ convolutional (conv) layer with $F_i$ kernels to reduce its number of channels to an intermediate value $F_i$.
The low-level feature map is then down-sampled to align with the shape of $g$. Next, the two resulting feature maps are combined by passing them to an element-wise summation operation followed by a ReLU function. A $1 \times 1$ conv layer with only one kernel is further applied to aggregate the information across all channels. After that, a sigmoid function is used to normalize the information and produce a coarse attention map, which is then resampled to match the shape of $x$. Finally, the fine-grained attention map $\alpha \in \mathbb{R}^{H_x \times W_x}$ is used to scale the feature map $x$.

In the first UNet of our proposed network (Fig.~\ref{fig:fig_attention_ResCUNeSt}), the attention gate takes a coarse feature map $D\_U_1$ from decoder and a low-level feature map $E\_U_1$ from the encoder as input and produces a filtered map $\hat{E}\_U_1$. The feature map $D\_U_1$ is then upsampled and concatenated with $\hat{E}\_U_1$ before fitting them to two successive  $3 \times 3$ conv layers followed by ReLU and Batch Norm. A similar mechanism is applied in the second UNet. However, in our design, $\hat{E}\_U_2$ is concatenated with not only $D\_U_2$ but also the coarse feature map $D\_U_1$ passed through the skip connections across the two UNets.

\subsection{Loss Function}
It is known that the problem of medical image segmentation poses an issue of imbalanced data, i.e., lesions or polyps are often small regions in an image. Therefore, we propose to employ  Tversky loss \cite{salehi2017tversky} to address the issue of data imbalance and achieve a much better tradeoff between precision and recall during training the networks. Assume that $P$ and $G$ are the predicted map taken after the softmax layer and the binary ground-truth, respectively, the Tversky loss is defined as follows:
\begin{equation}
T(\alpha, \beta, P, G) = \frac{\sum_{i=1}^{N}P_{i0}G_{i0}}{\sum_{i=1}^{N}P_{i0}G_{i0} + \alpha \sum_{i=1}^{N}P_{i0}G_{i1} + \beta \sum_{i=1}^{N}P_{i1}G_{i0}},
\end{equation}
where $N$ is the number of pixels in the ground-truth $G$; $P_{i0}$ is the probability that pixel $i$ belongs to a polyp, $P_{i1} = 1 - P_{i0}$ is the probability that pixel $i$ belongs to a non-polyp region; $G_{i0} = 1$ for a polyp pixel, $G_{i0} = 0$ for a non-polyp pixel and vice verse for $G_{i1}$; $\alpha$ and $\beta$ control the magnitude of penalties for false positives and false negatives, respectively.

An auxiliary Tversky loss is also applied to the first UNet to boost the gradient flow during training. Thus, the final loss function is:
\begin{equation}
L = T(\alpha, \beta, P_{U2}, G) + T_{aux}(\alpha, \beta, P_{U1}, G),
\end{equation}
where $P_{U1}$ and $P_{U2}$ are the output of the first and the second UNet, respectively.

\section{Experiments}
\label{sec:experiments}

\subsection{Datasets}
Several benchmark datasets are available for evaluating polyp segmentation models. The CVC-ClinincDB \cite{bernal2015wm} and ETIS-Larib \cite{silva2014toward} datasets are provided in the 2015 MICCAI automatic polyp detection sub-challenge. These datasets consist of frames extracted from colonoscopy videos, annotated by expert video endoscopists. CVC-ClinicDB has 612 images (384x288) extracted from 29 different video studies. ETIS-Larib has a total of 196 high-resolution images (1225x966). The CVC-ColonDB \cite{bernal2012towards} dataset is contributed by the Machine Vision Group (MVG) and contains 15 different polyps in 380 images (574x500). Finally, the Kvasir-SEG dataset \cite{jha2020kvasir}, publicized by the Simula Research Laboratory, includes 1000 polyp images with varying sizes.

To evaluate the effectiveness of each proposed component in our new architecture and compare the performance of our model across datasets over state-of-the-art approaches, we conduct experiments with different scenarios of using training and test data, as follows:
\begin{itemize}
\item \textbf{Scenario 1}: CVC-Colon and ETIS-Larib for training, CVC-Clinic for testing;
\item \textbf{Scenario 2}: CVC-Colon for training, CVC-Clinic for testing;
\item \textbf{Scenario 3}: CVC-ClinicDB for training, ETIS-Larib for testing. This is the combination used in the 2015 MICCAI sub-challenge;
\item \textbf{Scenario 4}: The Kvasir-SEG and CVC-Clinic datasets are merged, then split 80/10/10 for training, validation, and testing. The test sets are kept separate for evaluation on each source dataset. This is the combination proposed by PraNet \cite{fan2020pranet};
\item \textbf{Scenario 5}: 5-fold cross-validation on the CVC-Clinic dataset, which is split into five equal folds. Each run uses one fold for testing and four folds for training;
\item \textbf{Scenario 6}: 5-fold cross-validation on the Kvasir-SEG dataset, which is split into five equal folds. Each run uses one fold for testing and four folds for training.
\end{itemize}

\subsection{Experiment Settings}
We implement the proposed models using the PyTorch framework. A single training run takes approximately 12 hours using an NVIDIA GTX 2080 GPU. Weights pre-trained on ImageNet for ResNet and ResNeSt are used to initialize the respective backbones. The training process consists of two phases. The first phase trains the first UNet to convergence, and the second phase trains the entire coupled network model. Both phases use stochastic gradient descent (SGD) with a learning rate of $5.10^{-3}$, and a momentum of $0.9$. 

\subsection{Data Augmentation}
The aforementioned datasets are generally small compared to other computer vision datasets, as annotations require expert endoscopists. Thus, image augmentation is quite often used to help diversify training data. Our experiments follow the Augmentation-II strategy proposed in \cite{shin2018automatic}. Particularly, we apply the following transformations to every training image:
\begin{itemize}
\item Rotating the images by 90, 180, and 270 degrees, respectively;
\item Flipping the images both horizontally and vertically;
\item Resizing the images with four scale factors of 0.9, 1.1, 1.15, and 1.2, respectively;
\item Blurring the images with a kernel size of $5 \times 5$;
\item Brightening the images by using RandomBrightness in \cite{albumentations} with alpha = 1.5;
\item Darkening the images by using RandomContrast in \cite{albumentations} with alpha = 0.5.
\end{itemize}

\subsection{Evaluation Metrics}
We use the performance metrics listed in the MICCAI 2015 challenge \cite{miccai_polyp} to evaluate model performance: precision, recall, IoU (Jaccard score), and Dice score (F1). These are the most well-known measures for segmentation accuracy evaluation. Metrics are measured on the macro level: measurements are made on every image, then averaged on the whole dataset across all images.
\begin{align*}
Recall &= \frac{TP}{TP + FN}\\
Precision &= \frac{TP}{TP + FP}\\
Dice(P,G) &= \frac{2TP}{2TP + FP + FN}\\
IoU(P,G) &= \frac{TP}{TP + FP + FN}
\end{align*}
where $P$ represents the model's prediction, $G$ is the ground-truth, TP is true positives, FP is false positives, and FN is false negatives.

\section{Results and Discussion}
\label{sec:results}
\subsection{Ablation Study}
In this section, we measure the impact of each component in the proposed model. For the ablation study, we choose Scenario 1, i.e., CVC-Colon and ETIS-Larib are used for training, and CVC-Clinic is used for testing, due to two following reasons. Firstly, there are a number of options for the model's architecture. Hence, we should choose a combination with a small training dataset to speed up the evaluation process. The CVC-ColonDB dataset, including 380 images, seems insufficient to fit large backbones. The ETIS-Larib may be a good addition to the training dataset. Secondly, the training and test datasets are taken separately from different sources with different image properties and characteristics. This cross-dataset experiment setup is useful to evaluate the generalization capability of the model. Table~\ref{tab_colon_etis_clinic} shows the overall results of ablation study. It can be seen that AG-CUResNeSt-101 architecture obtained the best results over state-of-the-art models in terms of mDice and mIoU scores, the second-best in terms of precision, and is comparable to other models in terms of recall.

\begin{table}[ht!]
\centering
\caption{Performance metrics of model variants in Scenario 1, i.e., training on CVC-Colon and ETIS-Larib, testing on CVC-Clinic}
\begin{tabular}{ c | c c c c}
\hline
Method & mDice $\uparrow$      & mIoU $\uparrow$   & Recall $\uparrow$         & Precision $\uparrow$       \\
\hline
\hline
VGG16-UNet & 0.759 &    0.660 & 0.831 & 0.778 \\
Efficientnet-B0-UNet &  0.747 & 0.650 & 0.871 & 0.737 \\
Efficientnet-B1-UNet &  0.754 & 0.662 & \textbf{0.877} &    0.747 \\
Efficientnet-B2-UNet &  0.800 & 0.715 & 0.806 & 0.860 \\
Efficientnet-B3-UNet &  0.803 & 0.716 & 0.849 & 0.824 \\
Efficientnet-B4-UNet &  0.813 & 0.731 & 0.835 & 0.860 \\
\hline
\hline
ResNet34-UNet                  & 0.783          & 0.692          & 0.827          & 0.821           \\
ResNet50-UNet                  & 0.805          & 0.719          & 0.843          & 0.827           \\
ResNet101-UNet                 & 0.811          & 0.731          & 0.849          & 0.838           \\
ResNeSt50-UNet                 & 0.814          & 0.725          & 0.829          & 0.861           \\
ResNeSt101-UNet                & 0.816          & 0.739          & 0.813          & \textbf{0.888}           \\
\hline
\hline
Attention ResNet101-UNet          & 0.815          & 0.730          & \underline{0.863}          & 0.825           \\
Attention ResNeSt101-UNet       & \underline{0.829}          & \underline{0.749}          & 0.842          & 0.877           \\
\hline
\hline
Attention ResCUNet-101          & 0.820          & 0.736          & 0.859          & 0.838           \\
AG-CUResNeSt-101 & \textbf{0.833} & \textbf{0.754} & 0.840 & \underline{0.883}  \\
\hline
\end{tabular}
\label{tab_colon_etis_clinic}
\end{table}

\subsubsection{The Effectiveness of Backbone Networks}
We first evaluate the use of different encoder backbones. Several ResNet variants including ResNet34, ResNet50, ResNet101, ResNeSt50, and ResNeSt101 have been used. Besides, we also try other CNN architectures such as VGG16 and EfficientNet family from B0 to B4. The ResNeSt architecture uses channel-wise attention on separate branches to enrich their features. Table~\ref{tab_colon_etis_clinic} shows that ResNeSt101 gives the best overall performance. ResNet backbones generally perform better as size increases. ResNeSt101 also improves over ResNest50, but the improvement is quite marginal. ResNest101 achieves lower recall (to 0.813 from 0.829), suggesting that a larger ResNeSt such as ResNeSt152 would likely not yield significant improvements. ResNeSt101 backbone significantly outperforms VGG16 and yields slightly better results than EfficientNet-B4 in terms of mDice and mIoU scores.

\subsubsection{The Effectiveness of Attention Gate}
Next, we conduct experiments using two backbones, ResNet101 and ResNeSt101, integrated with the Attention Gate (AG) module. The integrated models are called Attention ResNet101-UNet and Attention ResUNeSt101-UNet, respectively. Table~\ref{tab_colon_etis_clinic} shows a considerable increase in mDice score when applying AG. More specifically, mDice for ResNet101-UNet increases from 0.811 to 0.815 when AG is added, while ResNeSt101-UNet increases from 0.816 to 0.829. We note that while the overall Dice score increases, adding AG causes a drop in precision score for both models. This is likely due to increased focus on potential polyp regions that had previously been ignored without AG. As attention gates bring more focus to these regions, the network can cover more polyps but is also more likely to make false predictions.

\subsubsection{The Effectiveness of coupled connections}
The Attention CUNet architecture adds one additional UNet, as well as skip connections across the two UNets. We denote the variant with ResNet backbone as Attention ResCUNet, and that with the ResNeSt backbone as AG-CUResNeSt. For each backbone, the mDice score increases by roughly $0.5\%$. AG-CUResNeSt achieves a mDice of 0.833 and a mIOU of 0.754, the best scores among models in Table~\ref{tab_colon_etis_clinic}. Both network size and the enrichment of semantic features play a factor in this improvement.

\subsection{Comparison to Existing Methods}
This section compares our proposed AG-CUResNeSt to several state-of-the-art models for polyp segmentation. From the previous ablation study, we select the best-performing ResNeSt101 backbone as the comparison model for this section. Therefore, the model is briefly called AG-CUResNeSt-101.

Six existing state-of-the-art models with publicly available source codes are used as baselines in the following evaluations, including: DDANet \cite{tomar2020ddanet}, ResUNet++ \cite{jha2019resunet++}, DoubleUNet \cite{jha2020doubleu}, HarDNet-MSEG \cite{huang2021hardnet}, ColonSegNet \cite{jha2021real} and PraNet \cite{fan2020pranet}. We re-use the original authors' reported metrics, and perform our own evaluations based on public codebases for un-tested datasets. Besides these baseline models, depending on each scenario, we also include other published results in literature for comparision where applicable.

\subsubsection{Cross-dataset Evaluation}
The following experiments evaluate the performance of AG-CUResNeSt-101 and previous state-of-the-art models when training and testing across different datasets, i.e., Scenario 1, Scenario 2, and Scenario 3. This setting implies that models need to generalize well to have good performance, as different polyp datasets have different image properties and feature distributions.

\begin{table}[ht!]
\centering
\caption{Performance metrics for ResUNet++, DoubleUNet, DDANet, ColonSegNet, HarDNet-MSEG, PraNet and AG-CUResNeSt-101 in Scenario 1, i.e., training on CVC-Colon and ETIS-Larib, testing on CVC-Clinic}
\begin{tabular}{ c | c c c c}
\hline

Method & mDice $\uparrow$      & mIoU $\uparrow$   & Recall $\uparrow$         & Precision $\uparrow$       \\
\hline
\hline

ResUNet++ \cite{jha2019resunet++} $\star$ & 0.406 & 0.302 & 0.481 & 0.496 \\

ColonSegNet \cite{jha2021real} $\star$ & 0.427 &    0.321 & 0.529 & 0.552 \\

DDANet \cite{tomar2020ddanet} $\star$ & 0.624 & 0.515 & 0.697 & 0.692 \\

DoubleUNet \cite{jha2020doubleu} $\star$ & 0.738 &  0.651 & 0.758 & 0.824 \\

HarDNet-MSEG \cite{huang2021hardnet} $\star$ & 0.765 &  0.681 & 0.774 & \underline{0.863} \\

PraNet \cite{fan2020pranet} $\star$ & \underline{0.779} &   \underline{0.689} & \underline{0.832} & 0.812 \\

\hline
\hline

AG-CUResNeSt-101 (Ours) & \textbf{0.833} & \textbf{0.754} & \textbf{0.840} & \textbf{0.883}  \\
\hline

\multicolumn{5}{l}{$\star$ indicates a model retrained with the original reported configurations.
}
\end{tabular}
\label{tab_colon_etis_clinic_2}
\end{table}

\begin{figure*}[ht!]
\centering
\includegraphics[width=350pt]{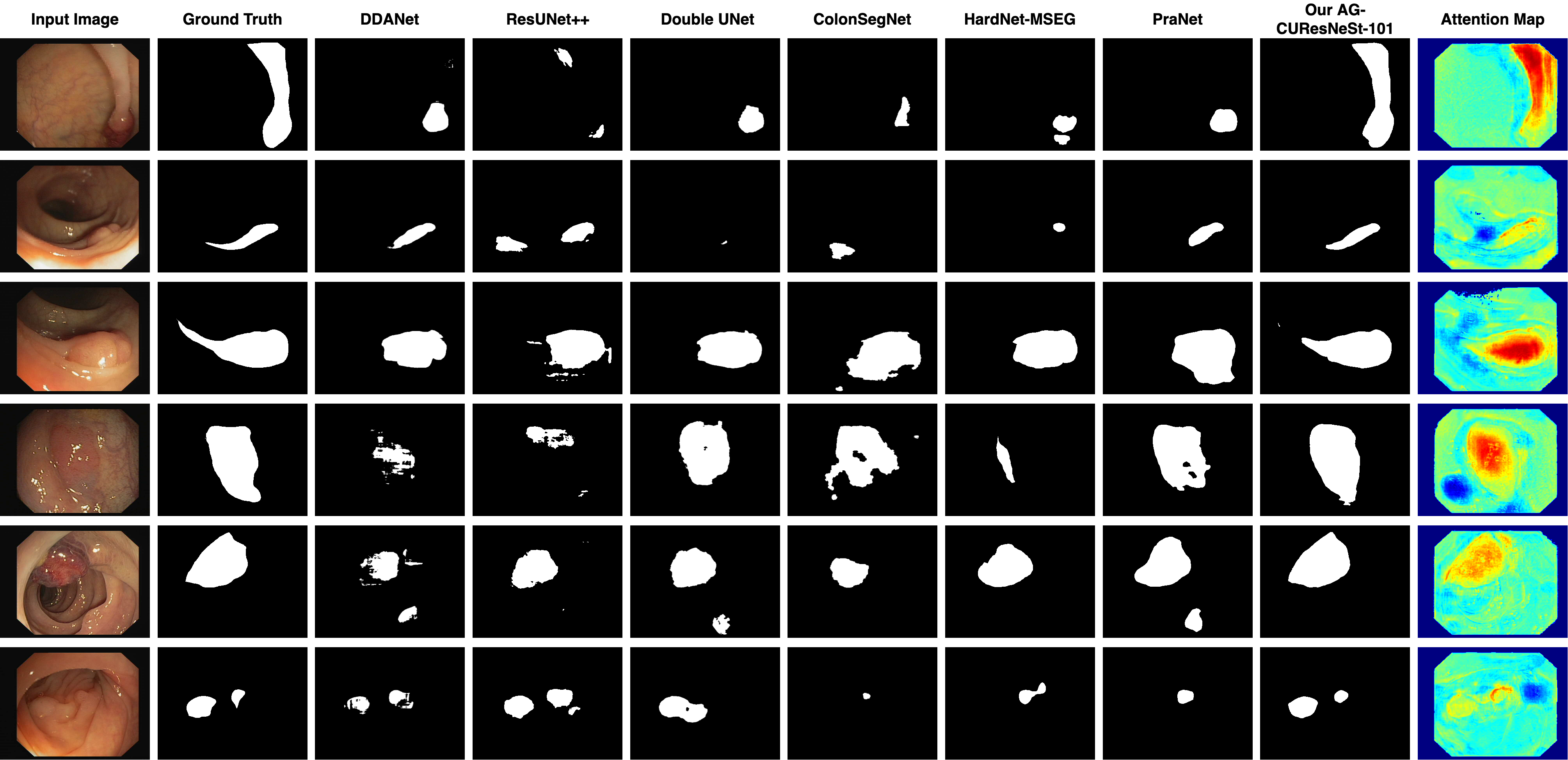}
\caption{Qualitative result comparison in Scenario 1, i.e., training on CVC-Colon and ETIS-Larib, testing on CVC-Clinic. From left to righ: input image, ground truth, outputs of DDANet, ResUNet++, Double UNet, ColonSegNet, HardNet-MSEG, PraNet, AG-CUResNeSt-101 (ours), and attention map in the last attention gate S9 in Fig.~\ref{fig:fig_attention_ResCUNeSt}. The red color in the attention map indicates the region where the model focus on.}
\label{fig_colon_etis_clinic}
\end{figure*}

We first compare the AG-CUResNeSt-101 with the baseline models on the train and test datasets used in our ablation study. Results are shown in Table~\ref{tab_colon_etis_clinic_2}. We find that AG-CUResNest-101 outperforms every baseline by relatively large margins. Against PraNet, the best-performing baseline, AG-CUResNeSt-101 achieves a $3.4\%$ improvement in Dice score and $5.5\%$ improvement in IoU score. These numbers are even larger compared to models such as ResUNet++ and ColonSegNet, which yield very low accuracy on the test set. Fig.~\ref{fig_colon_etis_clinic} shows qualitative results comparison between different methods on challenging images in Scenario 1. Our model can yield accurate segmentation in most of the cases, while all other baseline methods failed.

\begin{figure*}[ht!]
\centering
\includegraphics[width=350pt]{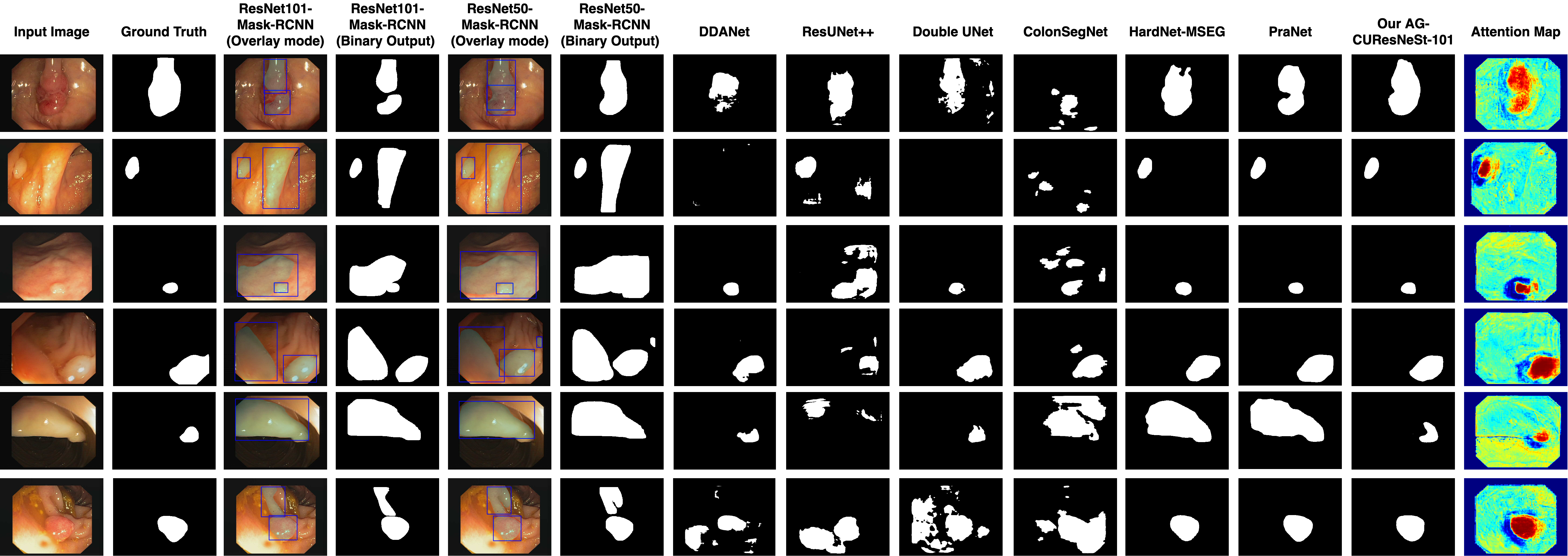}
\caption{Qualitative result comparison using CVC-Colon for training and CVC-Clinic for testing. From left to right: input image, ground truth, visualization of ResNet101-MaskR-CNN's output in overlay mode, binary output of ResNet101-MaskR-CNN, visualization of ResNet50-MaskR-CNN's output in overlay mode, binary output of ResNet50-MaskR-CNN, binary outputs of DDANet, ResUNet++, Double UNet, ColonSegNet, HardNet-MSEG, PraNet, our AG-CUResNeSt-101, and attention map in the last attention gate S9 in Fig.~\ref{fig:fig_attention_ResCUNeSt}. The red color in the attention map indicates the region where the model focus on.}
\label{fig_colon_clinic}
\end{figure*}

\begin{figure*}[ht!]
\centering
\includegraphics[width=300pt]{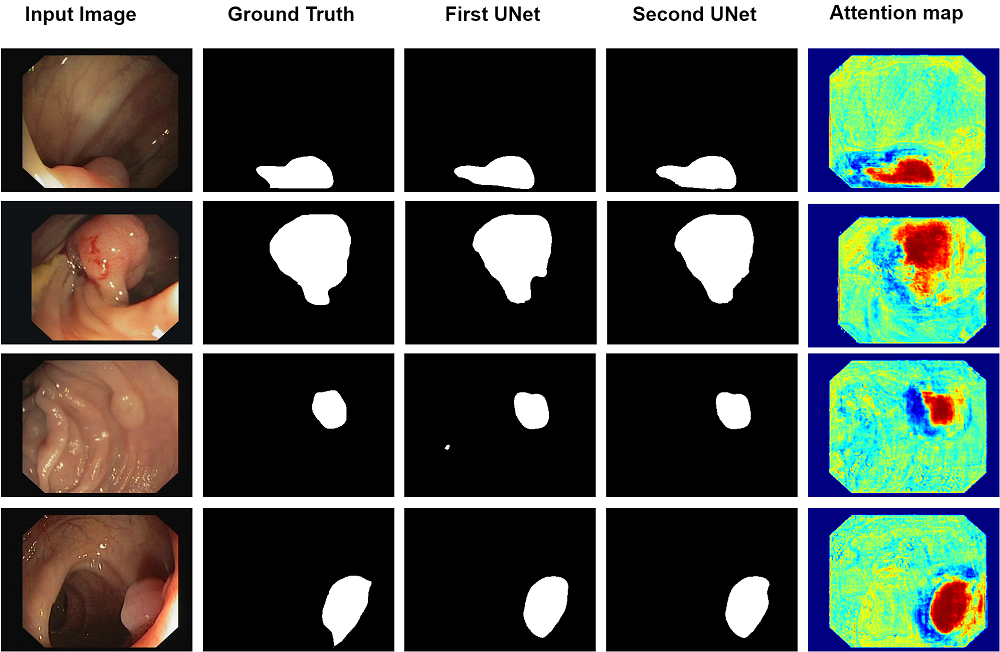}
\caption{The results of AG-CUResNeSt-101 on CVC-Clinic dataset. From left to right: input image, ground truth, output of the first UNet, output of the second UNet, and attention map in the last attention gate  S9 in Fig.~\ref{fig:fig_attention_ResCUNeSt}. The red areas in the attention map are high probability where polyps appear.}
\label{fig_compare_two_UNets}
\end{figure*}

We also compare AG-CUResNeSt-101 with Mask-RCNN \cite{qadir2019polyp} alongside the baseline models in Scenario 2, i.e., using CVC-Colon for training and CVC-Clinic for testing. We use the implementation of Mask-RCNN in the detectron2 project and train the model from scratch using the original paper's hyperparameters \cite{qadir2019polyp}. Table~\ref{tab_colon_clinic} shows that AG-CUResNeSt-101 outperforms Mask-RCNN by a large margin (by over $13\%$). PraNet is still the second-best baseline model, which is outperformed by AG-CUResNeSt-101 by $3.3\%$ in mDice score. Fig.~\ref{fig_colon_clinic} provides additional references for the output produced by each model. Notably, AG-CUResNeSt-101 seems capable of detecting both tiny and large polyps that occupy the whole image. We also compare the final output taken from the second UNet and the auxiliary output taken from the first UNet. Fig.~\ref{fig_compare_two_UNets} shows that the second UNet can correct some regions that the first UNet fails to predict.


\begin{table}[ht!]
\centering
\caption{Performance metrics for Mask-RCNN, ResUNet++, DoubleUNet, DDANet, ColonSegNet, HarDNet-MSEG, PraNet and AG-CUResNeSt in Scenario 2, i.e., using CVC-Colon for training, CVC-Clinic for testing}
\begin{tabular}{ c|c c c c}
\hline
Method & mDice $\uparrow$ & mIoU $\uparrow$ & Recall $\uparrow$ & Precision $\uparrow$  \\
\hline
\hline
ResNet50-Mask-RCNN \cite{qadir2019polyp}  $\star$ & 0.639    & 0.560        & 0.648  & 0.710      \\
ResNet101-Mask-RCNN \cite{qadir2019polyp} $\star$ & 0.641 & 0.565        & 0.646  & 0.725      \\

ResUNet++ \cite{jha2019resunet++} $\star$ & 0.339 & 0.247 & 0.380 & 0.484 \\

DoubleUNet \cite{jha2020doubleu} $\star$ & 0.441 &  0.375 & 0.423 & 0.639 \\

DDANet \cite{tomar2020ddanet} $\star$ & 0.476 & 0.370 & 0.501 & 0.644 \\

ColonSegNet \cite{jha2021real} $\star$ & 0.582 &    0.268 & 0.511 & 0.460 \\

HarDNet-MSEG \cite{huang2021hardnet} $\star$ & 0.721 &  0.633 & 0.744 & 0.818 \\

PraNet \cite{fan2020pranet} $\star$ & \underline{0.738} &   \underline{0.647} & \underline{0.751} & \textbf{0.832} \\
\hline
\hline
AG-CUResNeSt-101 (Ours)   & \textbf{0.771}    & \textbf{0.686}        & \textbf{0.793}  & \underline{0.830}  \\
\hline
\multicolumn{5}{l}{$\star$ indicates a model retrained with the original reported configurations.
}
\end{tabular}
\label{tab_colon_clinic}
\end{table}

\begin{table}[ht!]
\centering
\caption{Performance metrics for Mask-RCNN, Double UNet, ResUNet++, ColonSegNet, DDANet, PraNet, HarDNet-MSEG and AG-CUResNeSt in Scenario 3, i.e., using CVC-ClinicDB for training, ETIS-Larib for testing}
\begin{tabular}{c|c c c c}
\hline
Method & mDice $\uparrow$ & mIoU $\uparrow$ & Recall $\uparrow$ & Precision $\uparrow$  \\
\hline
\hline
ResNet50-Mask-RCNN \cite{qadir2019polyp}  $\star$    & 0.501   & 0.412         & 0.546  & 0.573      \\

ResNet101-Mask-RCNN \cite{qadir2019polyp}  $\star$   & 0.565   & 0.469         & 0.565  & 0.639      \\

DoubleUNet (BCE loss) \cite{jha2020doubleu} $\star$ & 0.482   & 0.400         & 0.713  & 0.475      \\

DoubleUNet (Dice loss) \cite{jha2020doubleu} $\star$ & 0.588   & 0.500         & 0.689  & 0.599      \\

ResUNet++ \cite{jha2019resunet++} $\star$ & 0.211 & 0.155 & 0.309 & 0.203 \\

ColonSegNet \cite{jha2021real} $\star$ & 0.217 &    0.110 & 0.654 & 0.144 \\

DDANet \cite{tomar2020ddanet} $\star$ & 0.400 & 0.313 & 0.507 & 0.464 \\

PraNet \cite{fan2020pranet} $\star$ & 0.631 &   0.555 & \textbf{0.762}  &    0.597\\

HarDNet-MSEG \cite{huang2021hardnet} $\star$ & 0.659 &  0.583 & 0.676 & \textbf{0.705} \\
\hline
\hline
AG-CUResNeSt-101 (Ours) & \textbf{0.701}  & \textbf{0.613}         & \underline{0.755}  & \underline{0.693}      \\
\hline
\multicolumn{5}{l}{$\star$ indicates a model retrained with the original reported configurations.
}
\end{tabular}
\label{tab_clinic_etis}
\end{table}

Next, we compare AG-CUResNeSt-101 with the baseline models in Scenario 3, i.e., using CVC-ClinicDB for training and ETIS-Larib for testing. Table~\ref{tab_clinic_etis} shows that AG-CUResNeSt-101 outperforms all other models in terms of mDice and mIoU, both by significant margins. HarDNet-MSEG is the second-best baseline in this scenario, achieving a Dice score of $0.659$. AG-CUResNeSt-101 outperforms HarDNet-MSEG by $4.2\%$ in Dice score and $3\%$ in IoU score. While AG-CUResNeSt-101 has slightly lower precision than HarDNet-MSEG, it achieves significantly higher recall.

\subsubsection{Intra-dataset Evaluation}
The following experiments compare AG-CUResNeSt-101 with several existing models when the training and test set are drawn from the same dataset, i.e., Scenario 4, 5, and 6.

Table~\ref{tab_kvasir_clinic} shows our evaluation results in Scenario 4, i.e., Kvasir-SEG and CVC-Clinic datasets are merged, then split 80/10/10 for training, validation, and testing. The proposed AG-CUResNeSt-101 is compared with the aforementioned baseline models along with reported metrics from SFA \cite{fang2019selective}, ResUNet-mod \cite{zhang2018road}, UNet \cite{ronneberger2015u} and UNet++ \cite{zhou2019unet++}. For the Kvasir-SEG test set, AG-CUResNeSt-101 outperforms the second-best PraNet by $0.4\%$ in mDice and $0.5\%$ in mIoU. For the CVC-ClinicDB test set, AG-CUResNeSt-101 is also the best performing model, exceeding PraNet by $1.9\%$ in mDice and $1.8\%$ in mIoU, respectively. We note that models such as DDANet and ColonSegNet perform much better in this scenario than the previous cross-dataset scenarios, implying that they may not be as robust to diverse test images as other models.

\begin{table} [ht!]
\centering
\caption{mDice and mIoU scores for models trained in Scenario 4 on the Kvasir-SEG and CVC-ClinicDB test sets}
\begin{tabular}{c|c c | c c}
\hline
\multirow{2}{*}{Method}               & \multicolumn{2}{c|}{Kvasir-SEG}  & \multicolumn{2}{c}{CVC-ClinicDB}   \\
\cline{2-3} \cline{4-5}
& mDice $\uparrow$ & mIoU $\uparrow$ & mDice $\uparrow$ & mIoU $\uparrow$  \\
\hline
\hline

UNet  \cite{ronneberger2015u}     & 0.818  & 0.746 & 0.823  & 0.755  \\
UNet++  \cite{zhou2019unet++}      & 0.821  & 0.743 & 0.794  & 0.729  \\
ResUNet-mod  \cite{zhang2018road}          & 0.791  & n/a   & 0.779  & n/a    \\
ResUNet++ \cite{jha2019resunet++}             & 0.813  & 0.793  & 0.796  & 0.796  \\
SFA \cite{fang2019selective}       & 0.723  & 0.611   & 0.700  & 0.607 \\

DDANet \cite{tomar2020ddanet} $\star$             & 0.758  & 0.658   & 0.761  & 0.668 \\

DoubleUNet \cite{jha2020doubleu} $\star$     & 0.781  & 0.700   & 0.791  & 0.730 \\

ColonSegNet \cite{jha2021real} $\star$     & 0.753 & 0.643   & 0.803  & 0.709 \\

HarDNet-MSEG \cite{huang2021hardnet} $\star$     & 0.877  & 0.807   & \underline{0.907}  & \underline{0.853} \\
PraNet \cite{fan2020pranet}       & \underline{0.898}  & \underline{0.840}   & 0.899  & 0.849 \\
\hline
\hline
AG-CUResNeSt-101 (Ours) & \textbf{0.902} & \textbf{0.845}  & \textbf{0.917}  & \textbf{0.867} \\
\hline
\multicolumn{5}{l}{$\star$ indicates a model retrained with the original reported configurations.
}
\end{tabular}
\label{tab_kvasir_clinic}
\end{table}

Next, we compare AG-CUResNeSt-101 against the baseline models alongside UNet \cite{ronneberger2015u} and MultiResUNet \cite{ibtehaz2020multiresunet} in Scenario 5, i.e., 5-fold cross-validation on the CVC-Clinic dataset. Results are shown in Table~\ref{tab_clinic}. Note that the authors of UNet and MultiResUNet only reported their IoU scores on this dataset. Regardless, we can see AG-CUResNeSt-101 shows significant improvement in this metric, outperforming the second-best PraNet by $1.3\%$ in mDice and $1.8\%$ in mIoU.

\begin{table}[ht!]
\centering
\caption{Performance metrics for UNet, MultiResUNet, ResUNet++, DoubleUNet, DDANet, ColonSegNet, HarDNet-MSEG, PraNet and AG-CUResNeSt-101 in Scenario 5, i.e., 5-fold cross-validation on the CVC-Clinic dataset}
\resizebox{\textwidth}{!}{%
\begin{tabular}{c|c c c c}
\hline
Method & mDice $\uparrow$ & mIoU $\uparrow$ & Recall $\uparrow$ & Precision $\uparrow$  \\
\hline
\hline
UNet \cite{ronneberger2015u}                   & -          & 0.792              & -          & -           \\
MultiResUNet \cite{ibtehaz2020multiresunet}& -          & 0.849              & -          & -           \\

ResUNet++ \cite{jha2019resunet++} $\star$ & $0.815 \pm 0.018$ & $0.736 \pm 0.017$ & $0.832\pm 0.018$ &  $0.830 \pm 0.020$ \\

DoubleUNet \cite{jha2020doubleu} $\star$ & $0.920 \pm 0.018$ & $0.866 \pm 0.025$ & $0.922 \pm 0.027$ & $0.928 \pm 0.017$ \\

DDANet \cite{tomar2020ddanet} $\star$ & $0.860 \pm 0.014$ & $0.786 \pm 0.017$ & $0.858 \pm 0.023$ & $0.892 \pm 0.014$ \\

ColonSegNet \cite{jha2021real} $\star$ & $0.817 \pm 0.020$ & $0.873 \pm 0.024$ & $0.926 \pm 0.025$ & $0.933 \pm 0.014$ \\

HarDNet-MSEG \cite{huang2021hardnet} $\star$ & $0.923 \pm 0.020$ & $0.873 \pm 0.024$ & $0.926 \pm 0.025$ & $0.933 \pm 0.014$ \\

PraNet \cite{fan2020pranet} $\star$ & \underline{$0.933 \pm 0.012$} & \underline{$0.884 \pm 0.015$} & \underline{$0.940 \pm 0.005$} & \underline{$0.937 \pm 0.016$} \\
\hline
\hline
AG-CUResNeSt-101 (Ours) & \textbf{0.946$\pm$0.01} & \textbf{0.902$\pm$0.015} & \textbf{0.953$\pm$0.013} & \textbf{0.944$\pm$0.009}  \\
\hline
\multicolumn{5}{l}{$\star$ indicates a model retrained with the original reported configurations.
}
\end{tabular}
}
\label{tab_clinic}
\end{table}

We also perform a comparison of AG-CUResNeSt-101 with the baseline models in Scenario 6, i.e., 5-fold cross-validation on the Kvasir-SEG dataset. Table~\ref{tab_kvasir} shows that the proposed model achieves the best mDice score, mIoU, recall, and precision. Specifically, AG-CUResNeSt-101 achieves an average Dice score of $0.912$, outperforming the second-place PraNet by $2.9\%$. Besides, metric scores across different folds demonstrate the stability of AG-CUResNeSt-101, with a slightly lower standard deviation than PraNet.

\begin{table}[ht!]
\centering
\caption{Performance metrics of UNet, ResUNet++, PraNet, DoubleUNet, DDANet, ColonSegNet, HarDNet-MSEG and AG-CUResNeSt-101 in Scenario 6, i.e., 5-fold cross-validation on the Kvasir-SEG dataset}
\resizebox{\textwidth}{!}{%
\begin{tabular}{c|c c c c}
\hline
Method & mDice $\uparrow$ & mIoU $\uparrow$ & Recall $\uparrow$ & Precision $\uparrow$  \\
\hline
\hline
UNet \cite{ronneberger2015u}                 & 0.708$\pm$0.017          & 0.602$\pm$0.01         & 0.805$\pm$0.014  & 0.716$\pm$0.02      \\
ResUNet++ \cite{jha2019resunet++}            & 0.780$\pm$0.01          & 0.681$\pm$0.008         & 0.834$\pm$0.01  & 0.799$\pm$0.01      \\

DoubleUNet \cite{jha2020doubleu} $\star$ & $0.879 \pm 0.018$ & $0.816 \pm 0.026$ & $0.902 \pm 0.027$ & $0.894 \pm 0.039$ \\

DDANet \cite{tomar2020ddanet} $\star$ & $0.860 \pm 0.005$ & $0.791 \pm 0.004$ & $0.876 \pm 0.015$ & $0.892 \pm 0.018$ \\

ColonSegNet \cite{jha2021real} $\star$ & $0.676 \pm 0.037$ & $0.557 \pm 0.040$ & $0.731 \pm 0.088$ & $0.730 \pm 0.080$ \\

HarDNet-MSEG \cite{huang2021hardnet} $\star$ & \underline{$0.889 \pm 0.011$} & \underline{$0.831 \pm 0.011$} & $0.892 \pm 0.015$ & \underline{$0.926 \pm 0.014$} \\
PraNet \cite{fan2020pranet} & 0.883$\pm$0.02          & 0.822$\pm$0.02         & \underline{0.897$\pm$0.02}  & 0.906$\pm$0.01      \\
\hline
\hline
AG-CUResNeSt-101 (Ours) & \textbf{0.912$\pm$0.01} & \textbf{0.860$\pm$0.011 }       & \textbf{0.923$\pm$0.009}  & \textbf{0.927$\pm$0.014}      \\
\hline
\multicolumn{5}{l}{$\star$ indicates a model retrained with the original reported configurations.
}
\end{tabular}
}
\label{tab_kvasir}
\end{table}

\begin{figure*}[!ht]
\centering
\includegraphics[width=\textwidth]{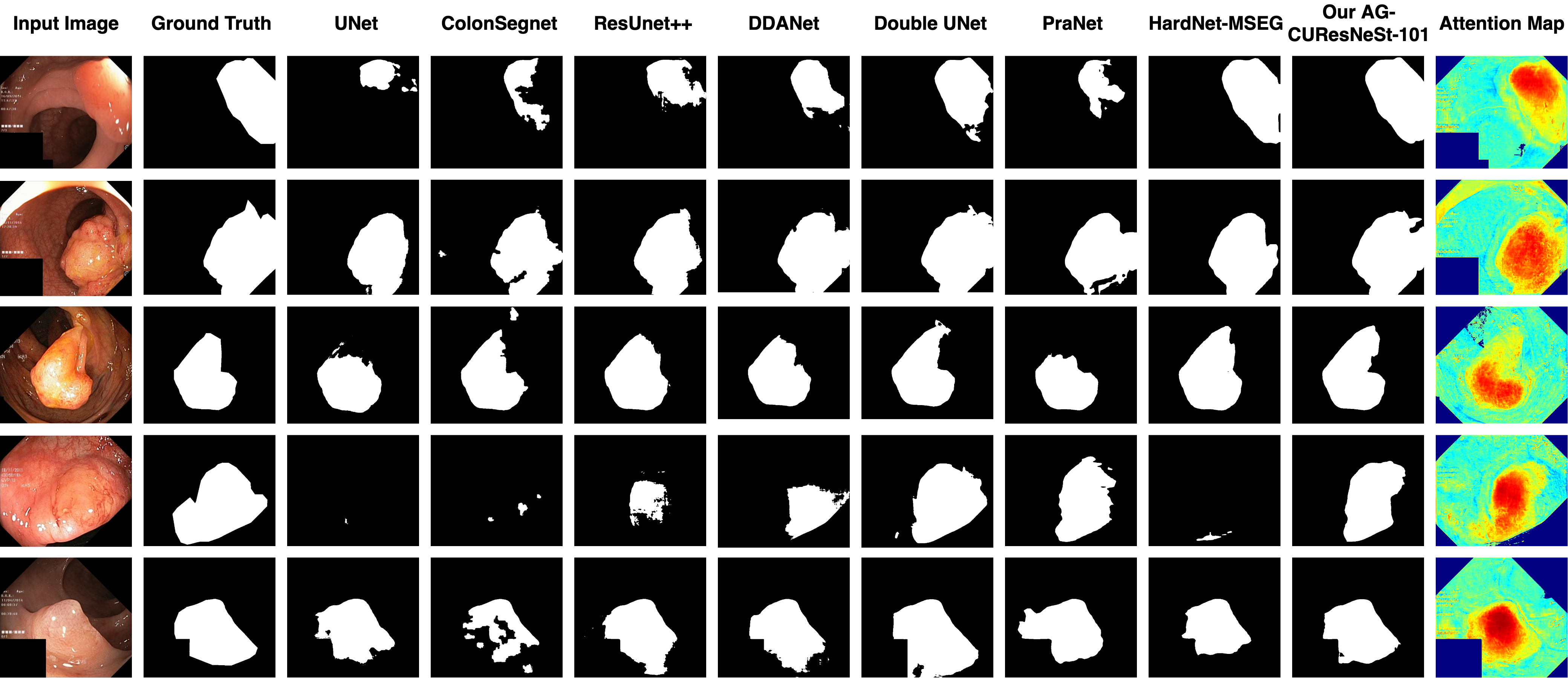}
\caption{Qualitative result comparison of different models trained in Scenario 6, i.e., 5-fold cross-validation on the Kvasir-SEG dataset.}
\label{fig_kvasir_results}
\end{figure*}

A qualitative comparison between different models is shown in Fig.~\ref{fig_kvasir_results}. In many cases, one can see that our model performs significantly better than other state-of-the-art methods. The lesion regions are well-segmented and delineated.

\begin{figure*}[!ht]
\centering
\includegraphics[height=0.4\textheight]{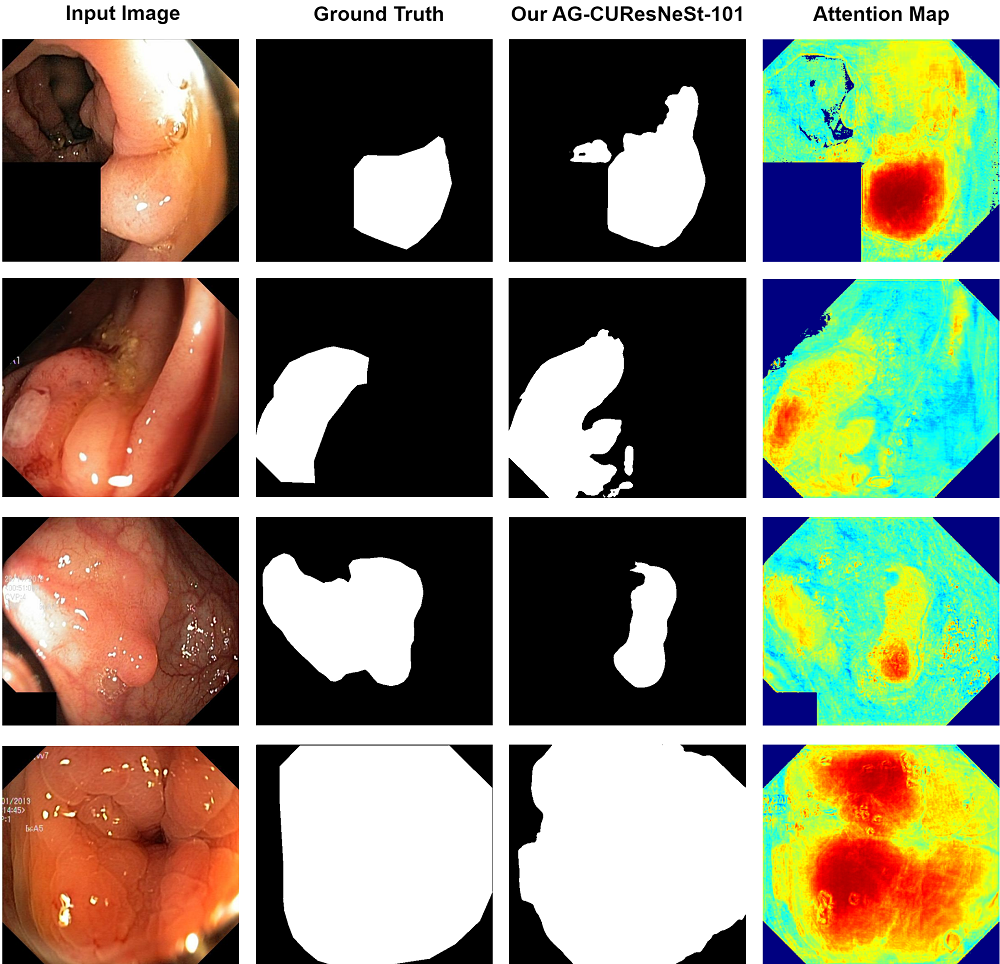}
\caption{Some failed cases of our model on the Kvasir-SEG dataset}
\label{fig_kvasir_failed_results}
\end{figure*}

Nevertheless, in some other case shown in Fig.~\ref{fig_kvasir_failed_results}, AG-CUResNeSt-101 is confused in estimating the attention maps, which lead to poor segmentation results. Usually, these imprecise predictions may occur when the input image contains very large polyps or colon mucosa folds, with many similar appearance characteristics as polyps. These are challenging cases for all the segmentation models and even junior endoscopists in practice.

\begin{figure} [h!]
\centering
\begin{subfigure}[b]{0.49\textwidth}
\centering
\includegraphics[width=\textwidth]{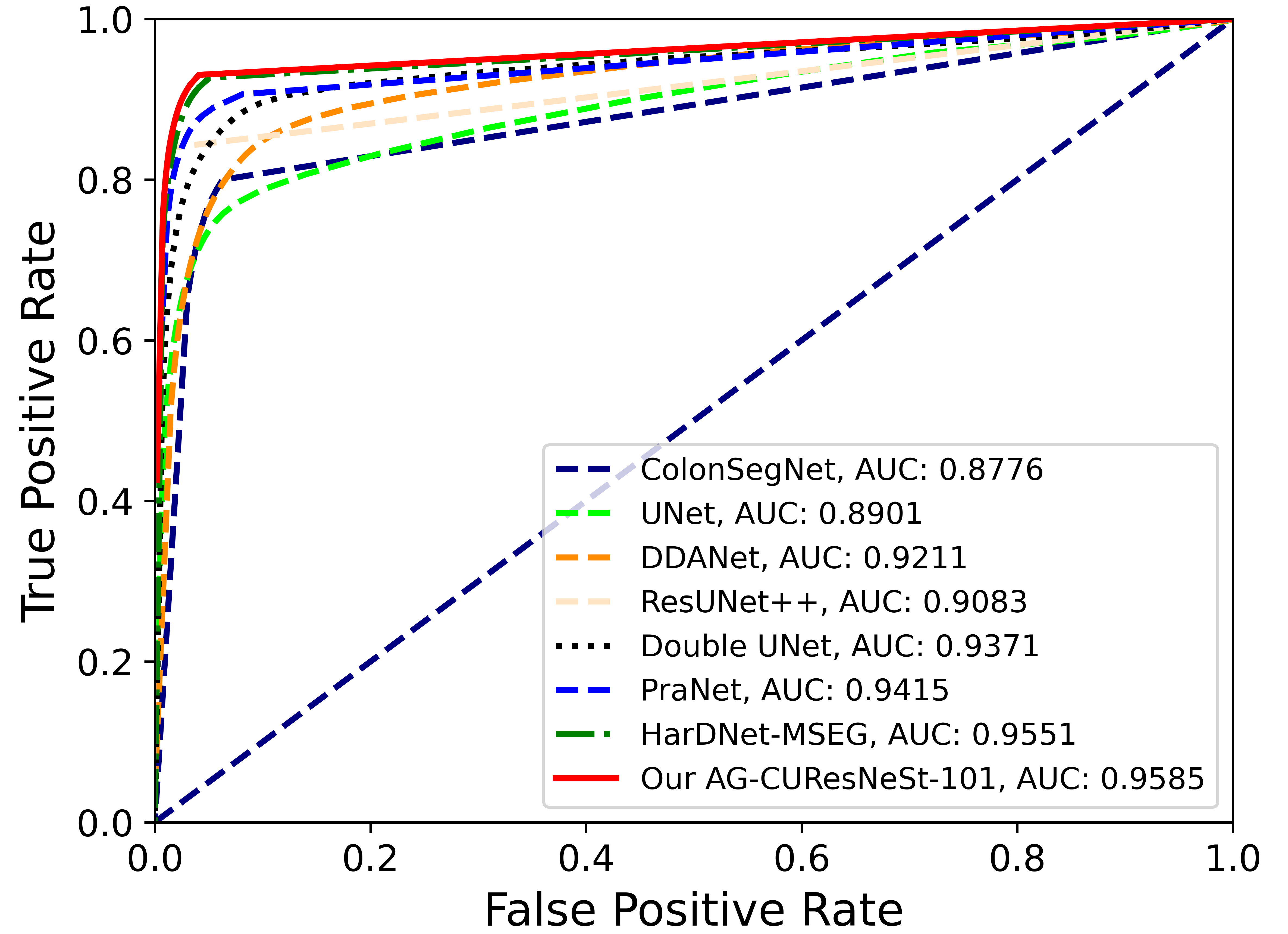}
\caption{ROC curves}
\end{subfigure}
\hfill
\begin{subfigure}[b]{0.49\textwidth}
\centering
\includegraphics[width=\textwidth]{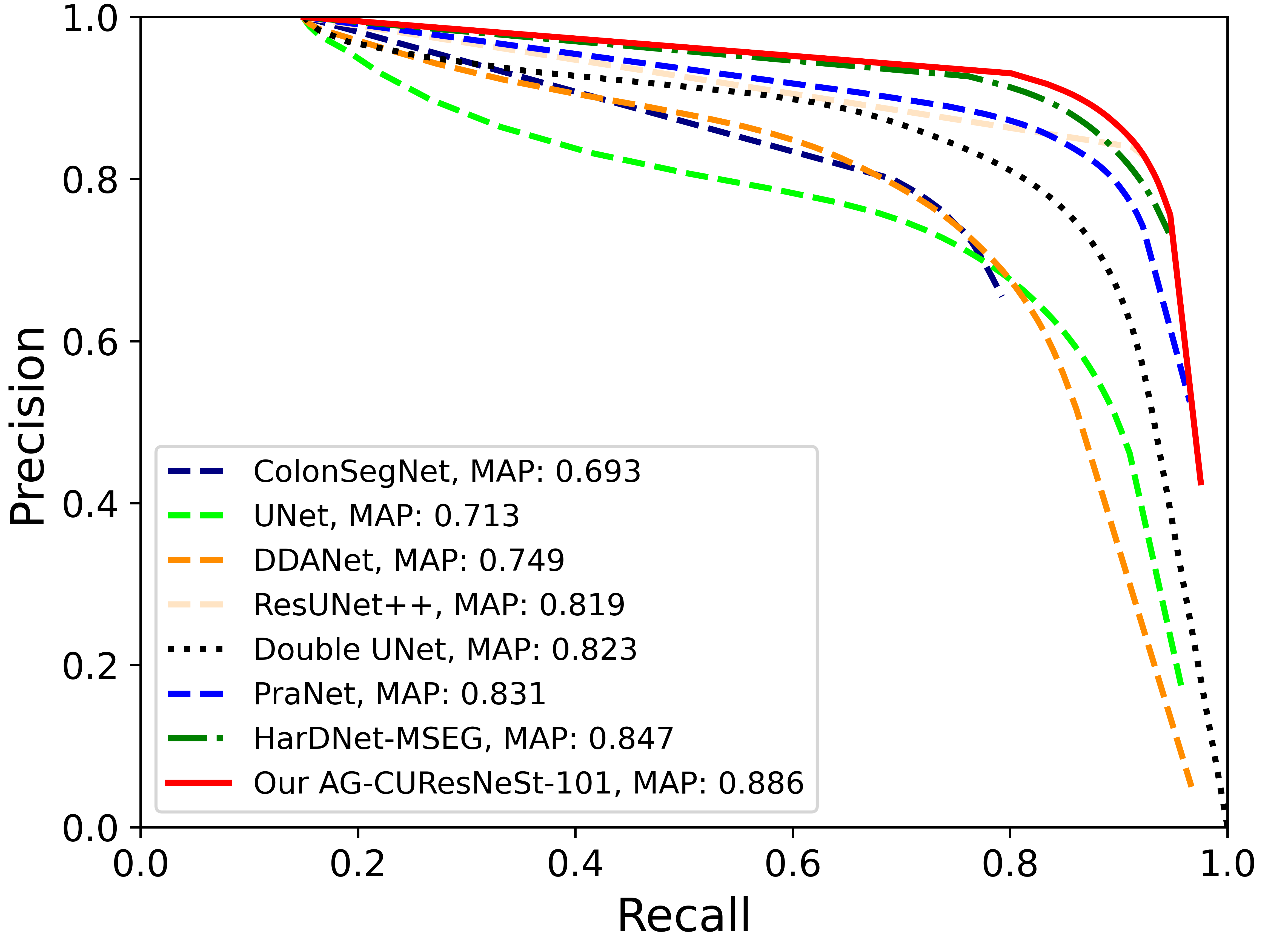}
\caption{PR curves}
\end{subfigure}
\caption{ROC curves and PR curves for AG-CUResNeSt-101 and other state-of-the-art models in Scenario 6, i.e., 5-fold cross-validation on the Kvasir-SEG dataset. All curves are averaged over five folds.}
\label{fig_kvasir_ROC}
\end{figure}

Fig.~\ref{fig_kvasir_ROC} shows the ROC curve and PR curve for each model in this experiment. Our AG-CUResNeSt-101 again reports the best AUC value of $0.9585$ and the best MAP value of $0.886$.

From the perspectives of endoscopists, the use of our proposed models is expected to support the training of junior staff in colon polyp detection. The improvements of our proposed model over state-of-the-art methods are especially helpful for inexperienced endoscopists in delineating lesions in challenging cases. Furthermore, it could be considered the possibility of setting up in clinical practice as a second-look tool or an assisting system during the procedure. The high accuracy of our novel models in benchmark datasets proposes a feasible solution to reduce the missing rate in real practice, which may have a significant impact on improving the quality of colorectal cancer screening strategy.

%


\subsection{Complexity and inference time comparison}

Finally, we compare the complexity and inference time of our proposed model and the state-of-the-art ones. The model complexity is measured as the number of floating operations (Flops), while the inference time per image is measured in seconds with corresponding Fps. We infer all the models using a machine with an NVIDIA GPU RTX 3090. Results are shown in Table \ref{tab_size}.

\begin{table}
\centering
\caption{Complexity and inference time comparison between our AG-CUResNeSt and other existing methods}
\begin{tabular}{c|c c c}
\hline

Model & GFlops & Inference time per image (sec) & Fps \\
\hline
\hline

ResUNet++ \cite{jha2019resunet++} & 283.42 & 0.056 & 17.9\\

ColonSegNet \cite{jha2021real} &  64.84 & 0.024 & 41.7\\

DDANet \cite{tomar2020ddanet} & 83.46 & 0.019 & 52.6\\

DoubleUNet \cite{jha2020doubleu} & 431.00 & 0.1 & 10 \\

HarDNet-MSEG \cite{huang2021hardnet} & 11.38 & 0.01 & 100 \\

PraNet \cite{fan2020pranet} & 13.11 & 0.014 & 71.4 \\

AG-CUResNeSt-101 & 273.40 & 0.052 & 19.2 \\
\hline
\end{tabular}
\label{tab_size}
\end{table}

We can see that the high accuracy achieved by our AG-CUResNeSt comes at a tradeoff in terms of floating-point operations. It is the third most complex model, behind ResUNet++ and DoubleUNet. DoubleUNet, in particular, has significantly more Flops than any other model at 431 GFlops. Meanwhile, ColonSegNet and DDANet are the most lightweight models, at fewer than 7 million parameters each. Nevertheless, these models also perform relatively poorly on cross-dataset testing experiments. Finally, HarDNet-MSEG and PraNet require the fewest floating operations despite their size, but their accuracy metrics still trails AG-CUResNeSt-101 significantly.
The complexity of AG-CUResNeSt-101 is relatively high since it doubles the UNet structure, but in return, our model outperforms other state-of-the-art methods in terms of accuracy. Note that, in clinical practice, each 1\% of adenoma detection rate increase is associated with a 3\% decrease in the risk of colon cancer. Therefore, it is worth it to trade off the model's complexity with its improvement of detection rate. Furthermore, our model still achieves 19.2 fps that is feasible for deployment on decent dedicated computing devices for real-time applications. This is particularly suitable for clinical practice in healthcare facilities with limited resources, especially in developing countries. The model is also suitable for retrospective studies for comparing the accuracy of an AI system with that of the original endoscopists in clinical practice.

\section{Conclusion}
\label{sec:conclude}
This paper has introduced a novel neural network architecture for polyp segmentation called AG-CUResNeSt. The architecture combines several components, namely ResNeSt, attention gates, and Coupled UNets, to improve performance. The proposed model is verified using extensive experiments and compared against several state-of-the-art methods on public benchmark datasets. Results show that AG-CUResNeSt consistently improves against all compared models, with a slight tradeoff in model complexity. Especially, our model achieves significantly better generalization capability in all cross-dataset experiments.

We hope that our proposal will provide a strong baseline for developing deep neural networks in medical image analysis, especially in colonoscopy. Our future research will focus on reducing the model size without sacrificing accuracy and conducting a deep study on failed cases to understand better the characteristics of the cases for further improvement in terms of segmentation accuracy. In addition, the generalizability issues will be addressed deeper by using more complicated approaches such as domain adaptation and domain generalization. 

\section{Acknowledgments}
This work is funded by Vingroup Innovation Foundation (VINIF) under project code VINIF.2020.DA17.

\bibliography{mybibfile}

\end{document}